\documentclass[]{spie}  
\usepackage[]{graphicx}
\usepackage{siunitx}
\usepackage{subcaption}	
\usepackage{sidecap}
\usepackage{wrapfig}

\DeclareSIUnit{\cmps}{\cm\per\second}
\DeclareSIUnit{\mps}{\meter\per\second}
\DeclareSIUnit{\kmps}{\kilo\meter\per\second}
\DeclareSIUnit{\micrometer}{$\mu$\meter}
\DeclareSIUnit{\foot}{'}
\DeclareSIUnit{\inch}{"}

\title{Development and construction of MAROON-X} 

\author{Andreas Seifahrt\supit{a}, Jacob L. Bean\supit{a}, Julian St\"urmer\supit{a},
Luke Gers\supit{b}, Deon S. Grobler\supit{b}, Tony Reed\supit{b}, and Damien J. Jones\supit{c} 
\skiplinehalf
\supit{a}University of Chicago, USA; \\
\supit{b}KiwiStar Optics, Callaghan Innovation, New Zealand;\\
\supit{c}Prime Optics, Australia 
}

\authorinfo{Further author information: (Send correspondence to A.S.)\\A.S.: E-mail: seifahrt@uchicago.edu, Telephone: 1 773 702 9877}

\begin{document} 
  \maketitle 

\begin{abstract}
We report on the development and construction of a new fiber-fed, red-optical, high-precision radial-velocity spectrograph for one of the twin 6.5m Magellan Telescopes in Chile. MAROON-X will be optimized to find and characterize rocky planets around nearby M dwarfs with an intrinsic per measurement noise floor below 1\,m\,s$^{-1}$. The instrument is based on a commercial echelle spectrograph customized for high stability and throughput. A microlens array based pupil slicer and double scrambler, as well as a rubidium-referenced etalon comb calibrator will turn this spectrograph into a high-precision radial-velocity machine. MAROON-X will undergo extensive lab tests in the second half of 2016.
\end{abstract}


\section{INTRODUCTION}
\label{sec:intro}  
The radial velocity (RV) method has been one of the most important observational techniques in the history of exoplanet science, and while it no longer drives the field, the technique will continue to be critical for making many of the most significant exoplanet discoveries anticipated over the next two decades through support of other experiments. For example, ground-based radial velocity measurements can deliver confirmation and mass measurement of candidate transiting planets, both from ground-based surveys and from space missions such as \textit{Kepler}, \textit{K2}, and \textit{TESS}. This is important because masses should be known for targets of follow-up atmospheric studies using \textit{HST}, \textit{Spitzer}, and \textit{JWST}. Knowing both the mass and radius of a planet constrains its bulk composition and surface gravity, which are crucial boundary conditions for the interpretation of spectra. Furthermore, populating the mass-radius diagram for exoplanets is currently a major topic in the field, especially in the regime of small planets \cite{gettel}.

In addition to the synergy with the transit technique, the radial velocity technique is also needed to support direct imaging efforts to study planets at intermediate to large separations. Simulations for \textit{WFIRST-AFTA} and a hypothetical flagship telescope with next-generation imaging capabilities (e.g., \textit{HabEx} or \textit{LUVOIR}) have shown that such missions would be much more successful if the planets were known ahead of time and had well-constrained masses and orbits \cite{stark,brown}.

It is thus not surprising that the field is seeing a variety of new RV spectrograph projects come to fruition. New ultra-stable instruments following the blueprint of HARPS have been commissioned or are being currently built, both for mid-size telescopes (e.g., HARPS-N/TNG, CARMENES/CAHA3.5m, SPIROU/CFHT, EPDS/WIYN) and large telescopes (e.g., ESPRESSO/VLT, IRD/Subaru, HPF/HET, G-CLEF/GMT) alike.

Our instrument, the \textsl{Magellan Advanced Radial velocity Observer Of Neighboring eXoplanets}, or MAROON-X, is following suit by providing the US community access to a precision RV instrument at a 6.5m telescope in the southern hemisphere. The scientific driver for MAROON-X is to detect the reflex motion of very low-mass stars due to the gravitational influence of an Earth-size planet orbiting in its habitable zone. For a 0.15\,M$_{\odot}$ star, the middle of the habitable zone is 0.055\,AU\cite{selsis}. A 1\,M$_{\oplus}$ planet at this distance around such a star would have an orbital period of 12.2\,d, and would impart a radial velocity signal of 1\,m\,s$^{-1}$ semi-amplitude assuming an edge-on orbit. This metric sets the desired level of precision for MAROON-X.

Beyond the required precision, we also aim for MAROON-X to have the reach necessary to take maximum advantage of the wealth of transiting planet candidates expected from the  \textit{TESS} mission \cite{sullivan}. In particular, we want to be able to obtain mass measurements for potentially habitable planets identified by TESS. Such planets would be ideal targets for \textit{JWST} spectroscopy to search for Earth-like environments \cite{deming}. The limiting factor for this experiment is the expectation for the distances to the nearest transiting, habitable-zone, Earth-size planets because these planets will be around stars that are farther away than the nearest such non-transiting planets. 

Based on the planet occurrence statistics from the \textit{Kepler} mission, \textit{TESS} is expected to find 20 planets with radii less than 2\,R$_{\oplus}$ and receiving between 0.2 and 2 times the Earth's insolation around mid M dwarfs that are brighter than V=16.5 \cite{dressing,sullivan}. Therefore, we aim to build an instrument that will enable on-sky radial velocity measurements with a precision of 1\,m\,s$^{-1}$ for mid M dwarfs as faint as V=16.5 in 30 minutes or less.

\begin{figure}[!bt]
\centering
\includegraphics[width=1.0\linewidth,clip]{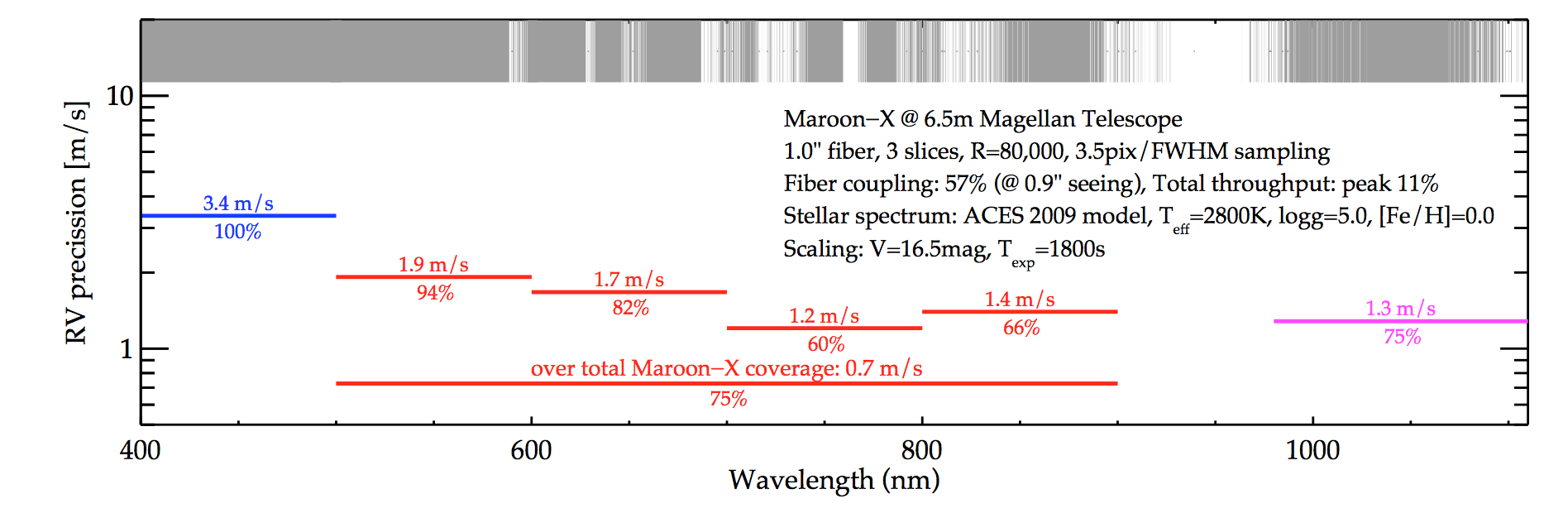}
\caption{\small{\textbf{M dwarf radial velocity information content as a function of wavelength interval.} We calculated the radial velocity information content in a T$_{\mathrm{eff}}$\,=\,2800\,K (M$_{\star}$\,$\approx$\,0.12\,M$_{\odot}$) model spectrum scaled to give a V-band magnitude of 16.5. We assumed a \SI{30}{\minute} exposure with a R\,=\,80,000 spectrograph on a 6.5m telescope. We used a complete system throughput model including the telescope, fiber feed, spectrograph, and detector over the MAROON-X bandpass (red lines). This model yields a peak throughput of 11\%, which is consistent with what has been achieved by existing optical spectrographs that deliver similar resolving power. We assumed a constant total system efficiency of 10\% outside this bandpass (blue optical: blue line; Y-band: purple line). The grey bar at the top indicates the strength of telluric absorption (white is fully opaque). Our calculations include masking off regions of strong telluric contamination. The number under each line gives the percentage of each spectral interval that is not contaminated. These results indicate that the optimum wavelength interval to observe M dwarfs for high-precision radial velocities is not the near-infrared, but the red part of the optical. The red optical outperforms the Y-band for stars down to 0.10\,M$_{\odot}$, and outperforms the J-, H-, and K-bands (not shown) down to the substellar limit. These findings have been verified using flux-calibrated empirical spectra of Barnard's Star taken with the HARPS, UVES, and CRIRES spectrographs, and they also confirm other previous studies based on model spectra\cite{reiners2010,rodler2011}.}}
\vspace{0mm}
\label{precisions}
\end{figure}

\section{INSTRUMENT CONCEPT}

Our concept for MAROON-X was driven by finding the most efficient solution to the technical requirements identified in our science case as outlined above. We have carried out detailed simulations to identify the optimum wavelength range to observe low-mass M dwarfs for radial velocity measurements to minimize the telescope aperture that is needed. Surprisingly, our conclusion is that the red part of the optical contains more radial velocity information than the near-infrared for stars down to masses of 0.10\,M$_{\odot}$ (T$_{\mathrm{eff}}$\,$\approx$\,2600\,K). See Figure \ref{precisions} for results of our simulations. The reason for this is that radial velocity measurements depend not just on the number of collected photons, but also on the spectral line density and the distribution of telluric lines in our own atmosphere. Although M dwarfs are brighter around 1\,$\mu$m, the very high line density at shorter wavelengths and the increased transparency of the Earths atmosphere at optical wavelength more than compensates for this. 

Beyond pure efficiency, there is also likely little to be gained in terms of reducing the influence of stellar activity on radial velocity measurements from going to longer wavelengths. Simulations of star spots on M dwarfs suggest that there is no further reduction in radial velocity jitter beyond \SI{1}{\micrometer} due to the reduced contrast between the normal stellar surface and the spots\cite{reiners2010}. Also, there is the possibility that the increased sensitivity of stellar lines at longer wavelengths to Zeeman splitting means that near-infrared spectra of M dwarfs could exhibit \textit{higher} jitter due to activity than optical spectra, which is the opposite of the usual assumption\cite{reiners2013}. Also, late M dwarfs are often faster rotators\cite{reinersbasri}. The detrimental influence of stellar rotation on radial velocity precision is a function of wavelength, again favoring the red optical over infrared wavelengths.

All this taken together suggests that a high-resolution spectrograph operating at red optical wavelengths is a good option for high-precision radial velocity measurements of very low-mass stars. We will observe at wavelengths of maximum radial velocity measurement efficiency for these stars, and we don't expect to be at a disadvantage in terms of activity-related jitter compared to other instruments.

Our calculations have shown that the intrinsic RV content of our targets is limited to about \SI{0.7}{\mps} for a V=16.5 M dwarf in a \SI{30}{\minute} exposure. Our instrumental stability budget must be under \SI{0.7}{\mps} as well in order to achieve our overarching goal of \SI{1}{\mps} on-sky performance.

Stabilizing a spectrograph to below \SI{1}{\mps} over long timescales requires a wide range of measures. The optical design has to eliminate moving parts and must deliver the needed spectral bandpass in one fixed setting. The spectrograph has to be mechanically and thermally decoupled from its environment to maintain pressure and temperature stability of \SI{\leq e-3}{\milli\bar} and \SI{\leq1}{\milli\kelvin} at the echelle grating. Therefore, it must be mounted in a vacuum chamber that is in turn housed in a temperature controlled enclosure. The instrument also must be fed with fiber optics for mechanical separation from the telescope and to stabilize the light injection both in the near field (light distribution in the slit) as well as the far field (light distribution in the pupil will change the PSF of the spectrograph). Additionally a fiber shaker is required to suppress modal noise and reduce speckle effects. Even at this extreme level of stability, RV drifts are expected, and simultaneous reference spectra with a dense comb spectrum are required to track and correct the long term zero-point of the spectrograph.

The main components and features that allow MAROON-X to achieve its technical specifications are thus:
\begin{itemize}
\item An echelle spectrograph with high throughout and thermo-mechanical stability (see Section\,\ref{kiwispec}).
\item Fiber feed with low focal-ratio-degradation and high scrambling efficiency (see Section\,\ref{fiber}).
\item Pupil slicer and image scrambler to match the slit-resolution product to the A$\Omega$ product of a 6.5\,m telescope and increase the near- and far-field stability of the light injection (see Section\,\ref{fiber}).
\item Fiber agitator to suppress modal noise and speckle effects.
\item A Fabry-P\'erot etalon as a wavelength calibrator capable of delivering \SI{<10}{\cmps} stability over timescales of minutes to months and a high information density to measure and correct for instrumental drifts (see Section\,\ref{etalon}). 
\item Environmental control with a temperature stability of the air surrounding the spectrograph of $dT$\,\SI{<10}{\milli\kelvin}. (see Section\,\ref{chamber})
\end{itemize}

\section{PROJECT OUTLINE}\label{outline}

Our project execution can be considered somewhat unusual in at least two aspects, owing to the limited available manpower and the challenging funding environment for a project of this size. For one, we decided to outsource the detailed technical design and construction of the core spectrometer to KiwiStar Optics, a subsidiary of Callaghan Innovation, a New Zealand government-owned Crown entity. 

KiwiStar Optics has developed an implementation of a versatile echelle spectrograph with 100\,mm beam diameter, 1:3 pupil compression and multiple camera arms, dubbed ``KiwiSpec R4-100'' \cite{barnes}. Our trade study has shown that this concept is very well suited for MAROON-X after some major modifications over their first prototype\cite{gibson} and custom input optics to match it to a 6.5\,m telescope. 

In answer to funding realities, we split the project in three phases. In phase 1, which we are now nearing completion, we modified the design of KiwiSpec R4-100 to match our requirements for throughput and stability. We held a PDR for the core spectrograph in June 2014 and ordered a two-arm version of KiwiSpec R4-100 (with only the blue arm executed) in June 2015. An internal FDR and kick-off meeting was held in August 2015 in Wellington, New Zealand. As of this writing all of the optics have been procured and delivered, the vacuum chamber has been completed, and integration is underway. We expect delivery of the spectrograph to Chicago in September 2016. 

In lieu of a large format (4k$\times$4k) detector and custom cryo- and readout systems we purchased only an off-the-shelf detector system with a smaller (2k$\times$2k) CCD for initial lab testing (see Section\,\ref{detector}). 

In  phase 1 we also developed the pupil slicer and image scrambler (see Section\,\ref{fiber}), a Fabry-P\'erot etalon as the wavelength calibrator system (see Section\,\ref{etalon}), characterized the fiber properties of octagonal and rectangular fibers (see Section\,\ref{fiber}), and built an actively controlled environmental enclosure to house the spectrograph (see Section\,\ref{chamber}).

In phase 2, which we hope to start in early 2017, we will complete the spectrograph with the red arm and build the two large-format science grade detector systems. Finally, in phase 3 we will build the telescope front end (ADC, guiding system, tip-tilt fiber feed) and commission the instrument at Magellan.

\section{KIWISPEC R4-100 as the core of MAROON-X}\label{kiwispec}
\subsection{Optics}

\begin{figure}[!b]
\centering
\includegraphics[width=1.0\linewidth,clip]{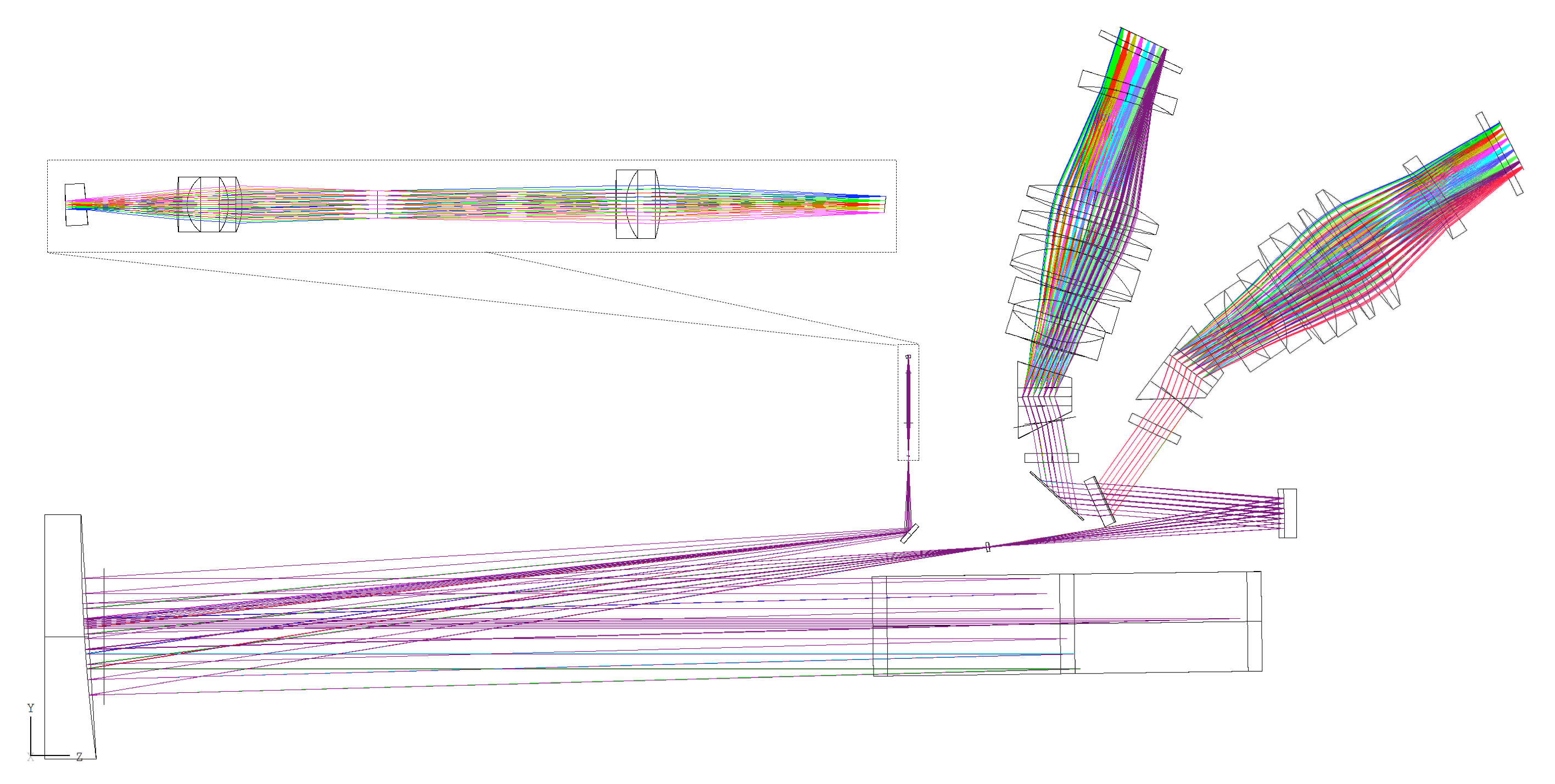}
\caption{\textbf{Zemax raytrace of the KiwiSpec R4-100 spectrograph for MAROON-X}. The blue arm (to the right) covers 500--670\,nm, the red arm (top) covers 650--900\,nm. The insert shows the telecentric input relay optics which convert the $f/5$ input to the $f/10$ accepted by the spectrograph. A wedge prism at the fiber exit introduces a field-dependent focus shift as part of the aberration control of the camera optics.}
\vspace{0mm}
\label{Kiwispec_raytrace}
\end{figure}

In our implementation for MAROON-X, the KiwiSpec R4-100 delivers a resolving power of R$\approx$80,000 for a \SI{100}{\micrometer} wide pseudo-slit at $f/10$ with 3.5\,pixel sampling across a wide wavelength range of 500 -- 900\,nm spread over two camera arms. A Zemax layout of the spectrograph is shown in Figure\,\ref{Kiwispec_raytrace}. 

\begin{figure}[!t]
\centering
\vspace{5mm}
\includegraphics[trim={0cm 0cm 0cm 0cm},width=0.55\linewidth,clip]{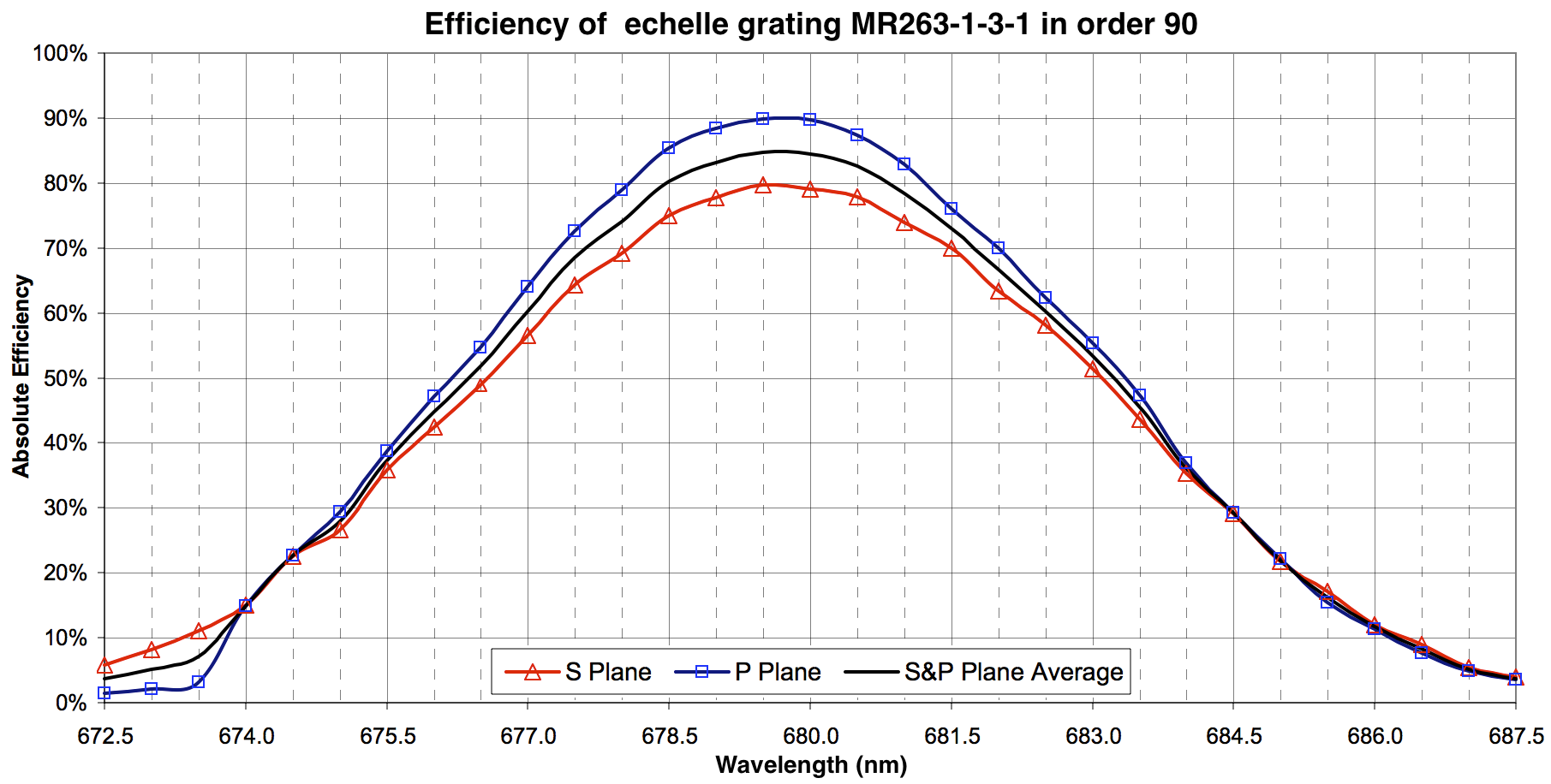}
\includegraphics[trim={20mm 36mm 25mm 20mm},width=0.44\linewidth,clip]{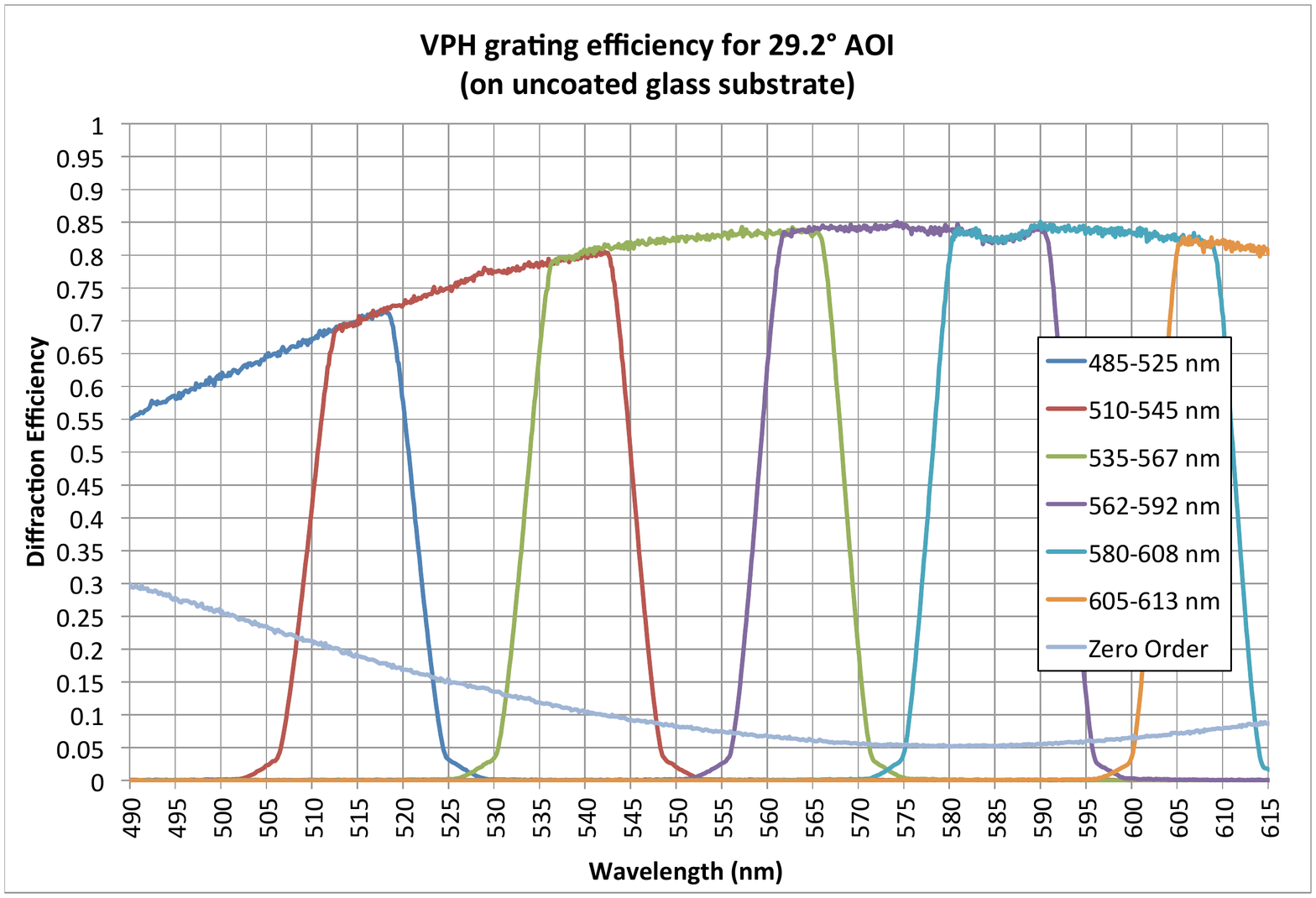}
\vspace{3mm}
\caption{\textbf{Measured efficiency of the echelle grating (left) and the blue VPH grating (right)}. The echelle grating was measured at Richardson Gratings. Shown is order 90. Measurements of the blue VPH grating by KiwiStar Optics (right) match the measurements by KOSI. The Fresnel losses on the uncoated glass substrate have not been corrected in this plot. In both cases peak efficiencies of 85\% were recorded.}
\label{dispersers}
\end{figure}

The main disperser is a Richardson Gratings R4 echelle with a nominal blaze of \si{\ang{76}} and a line frequency of 31.6\,g/mm. Our echelle was replicated from master MR263 on a \SI{116x420}{mm} Zerodur substrate with a \SI{102x408}{mm} ruled area and coated with protected silver. Measurements at Richards Gratings (see Figure\,\ref{dispersers}) show peak efficiencies of up to 85\% at blaze. The grating was remeasured in all orders from 400--1020\,nm at KiwiStar Optics, confirming the excellent efficiency. From the blaze maxima, a blaze angle of \SI[separate-uncertainty]{75.29 \pm 0.25}{\degree} was determined, slightly lower than specified.

The collimator mirror, pupil transfer mirror, and both fold mirrors are made from Zerodur class 1 and coated with protected enhanced silver. 

\begin{SCfigure}[][!b]
\centering
\vspace{0mm}
\includegraphics[trim={0cm 0cm 0cm 0cm},width=0.65\linewidth,clip]{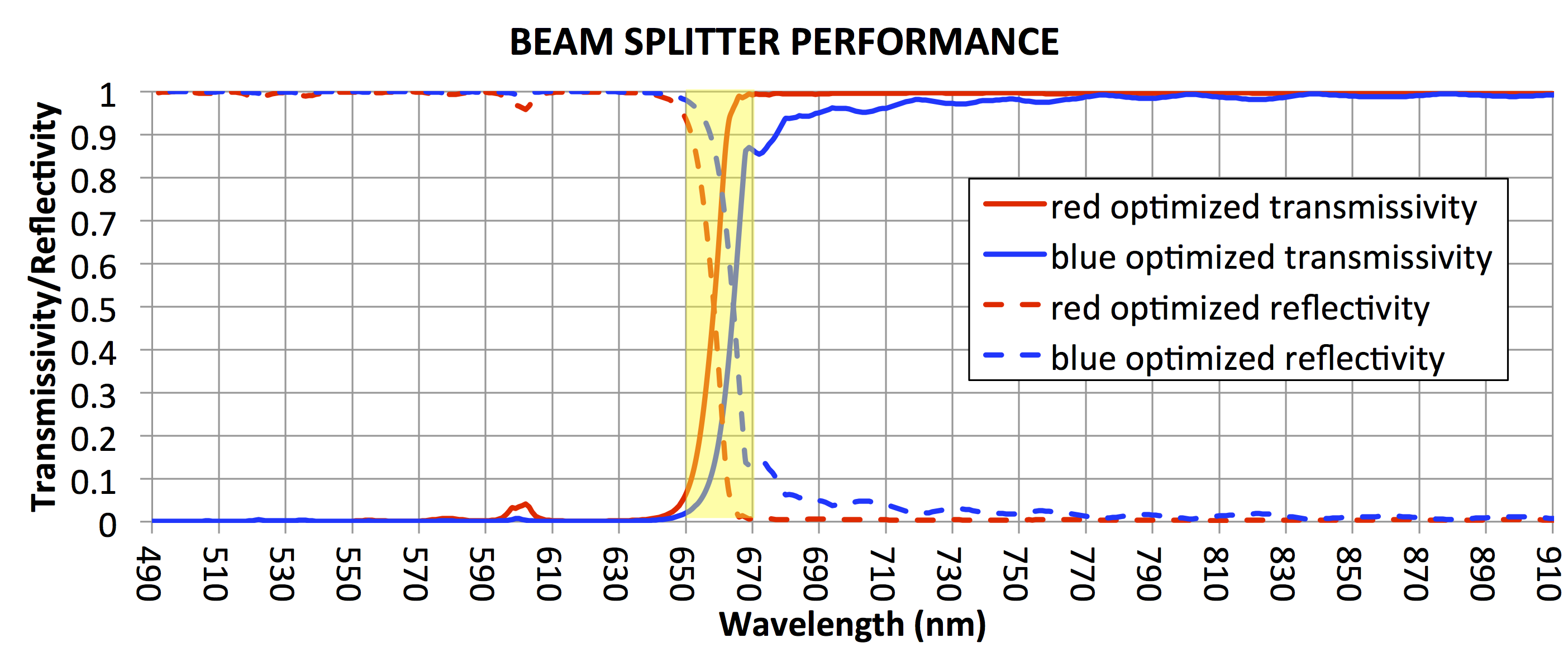}
\vspace{0mm}
\caption{\textbf{Theoretical performance of the dichroic beam splitter}. The minimum and maximum of the expected performance are shown as ``blue optimized'' and ``red optimized'' case, respectively. The yellow shaded region shows the order-overlap between the two spectrograph arms.}
\label{cascade}
\end{SCfigure}

A dichroic beam splitter separates the dispersed light into the two camera arms. The crossover wavelength is 660\,nm. Theoretical coating performance based on data from Cascade Optical Corp. is shown in Figure\,\ref{cascade}. The range of transmission and reflection efficiencies (including the BBAR coating on the backside) favor either the blue or the red wavelength channel. We have expressed a preference for the ``red optimized'' case, but will likely see an as-built performance intermediate between the two cases.

Two VPH cross-disperser grisms and camera arms cover 500--670\,nm and 650--900\,nm, respectively, with a considerable order overlap to minimize the losses on the slope of the dichroic. The blue VPH grating was produced by KOSI. Its line frequency is 1652.4\,l/mm and the AOI is \ang{29.2}. The peak efficiency was measured to be 85\% (see Figure\,\ref{dispersers}). Two prisms will be bonded onto the VPH grating that have BBAR coatings on the air-glass surfaces. We thus expect the efficiency of the VPH grism assembly to be slightly higher than the one measured for the bare grating. 

Our cameras have a focal length of $f$\SI{=174}{mm} and hence a moderate $f/5.22$. The large height of the slit image (2.7\,mm at $f/10$) in combination with the large field angles and the anamorphism of the VPH grism makes achieving a homogeneous optical quality over the whole field and for all wavelengths challenging and required a compromise between number of optical elements and requirements towards the PSF size and homogeneity. Our efforts resulted in two 9-element cameras, each with two aspheric surfaces. The last element of each camera is a plano-cylindric field flattener lens that serves as the dewar window for the detector. Part of both cameras is a small wedge prism at the input slit (directly bonded onto the fiber slit plate, see below), that introduces a field dependent focus shift, which slightly eases the aberration control in the main camera optics. 

\begin{figure}[!t]
\centering
\vspace{5mm}
\includegraphics[trim={1cm 0cm 1cm 0cm},width=0.49\linewidth,clip]{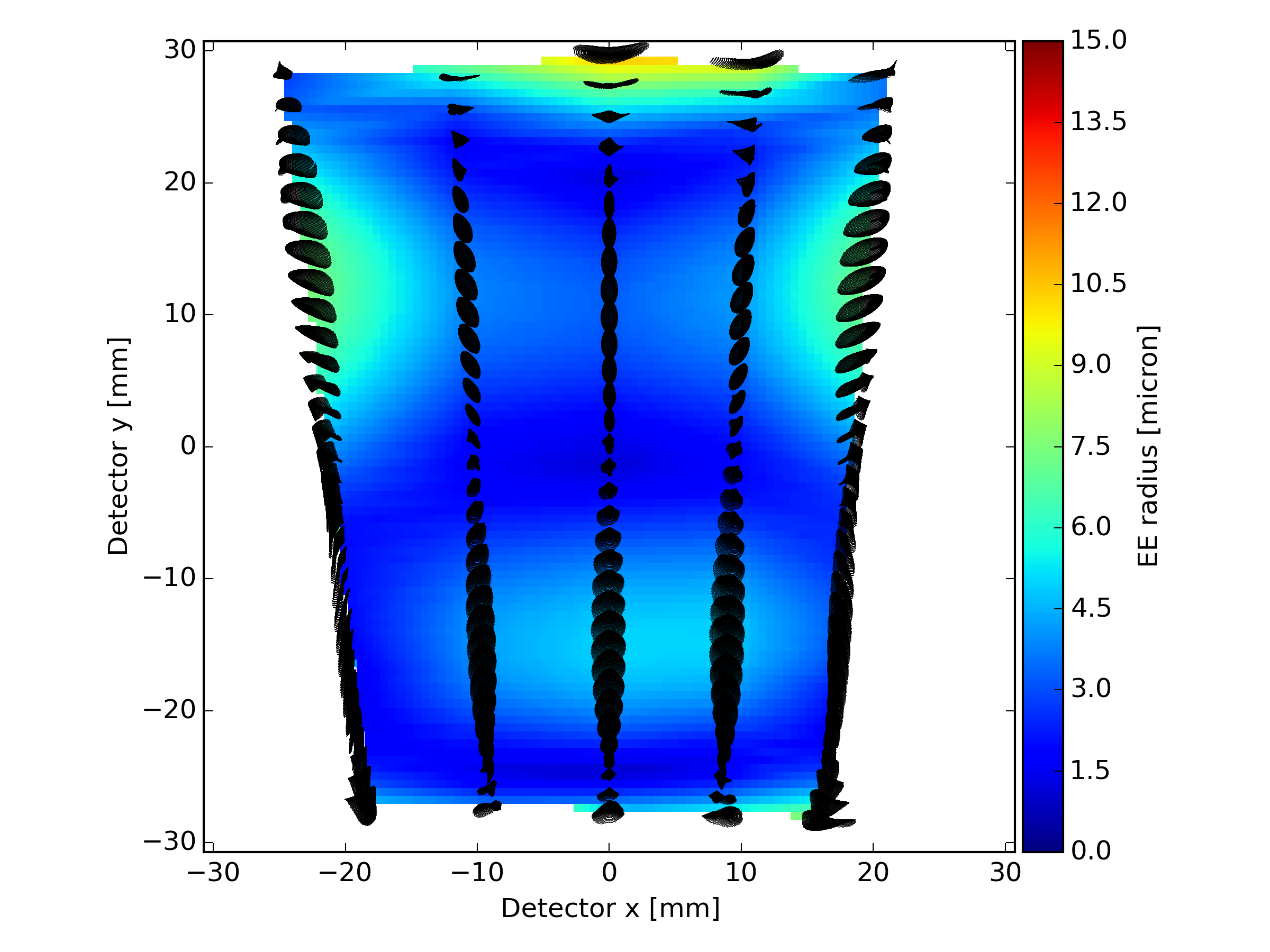}
\includegraphics[trim={1cm 0cm 1cm 0cm},width=0.49\linewidth,clip]{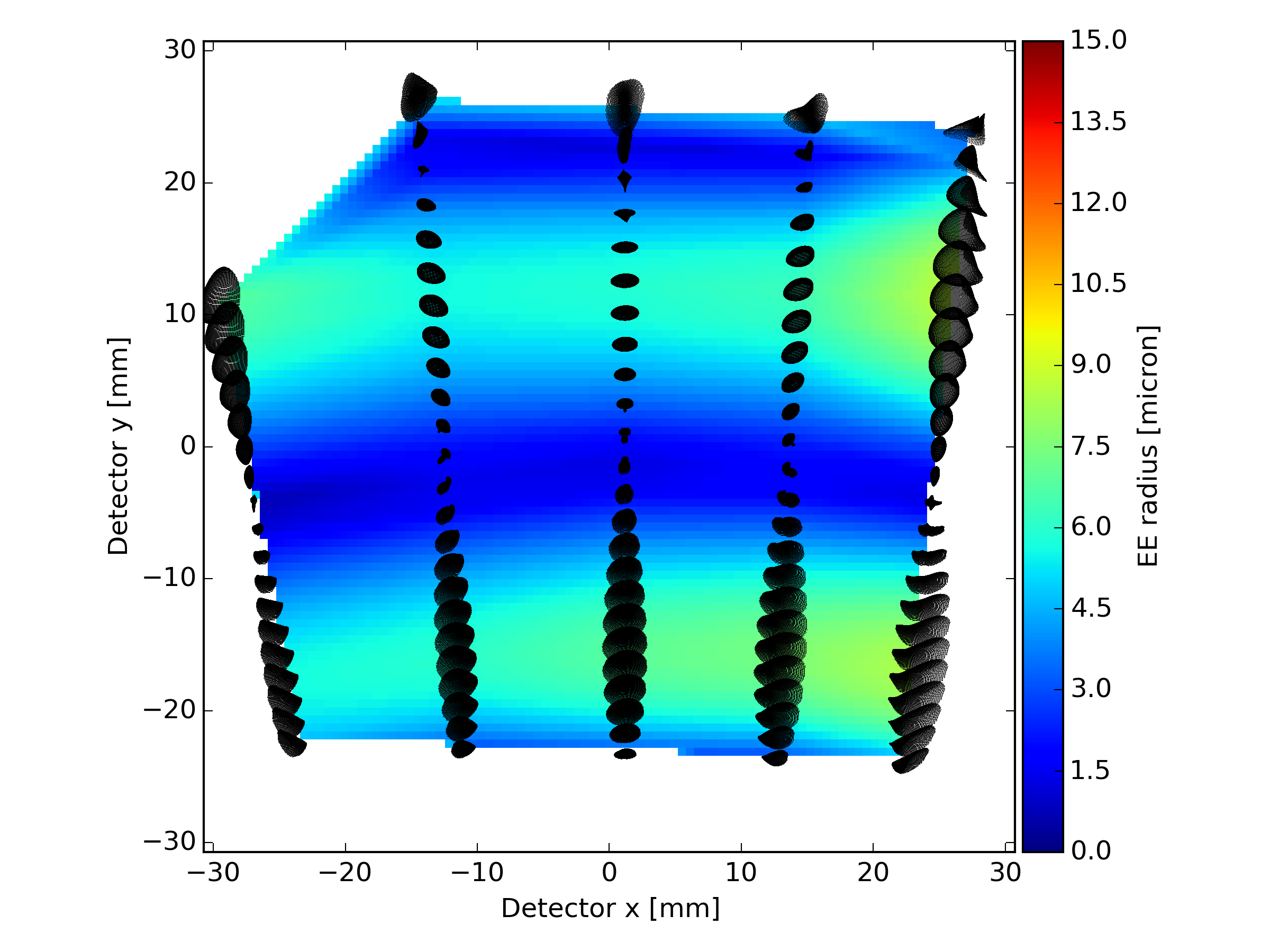}
\vspace{3mm}
\caption{\textbf{PSF simulations for the blue (left) and red (right) camera arm}. The half-width of the 80\% encircled-energy in the dispersion direction is shown as the color-coded map. For reference, one detector element is \SI{15}{\micrometer} and the average slit sampling is 3.5\,pixel or \SI{53}{\micrometer}.}
\label{Kiwispec_psf}
\end{figure}

We achieve typical PSF widths of \SI{\leq 15}{\micrometer} (80\% EE) in the dispersion direction and \SI{\leq 23}{\micrometer} (80\% EE) in the spatial direction at the center of the field (see Figure\,\ref{Kiwispec_psf}). Towards the edges of the field (the calibration and sky fibers) some regions in the spectrum, particularly at the edges of the FSR do exceed these limits. 

Most gratifyingly, the PSF shape in the dispersion direction is very symmetric, as is the change of the PSF shape over a given order. This greatly reduces RV shifts induced by varying pupil illuminations (e.g. from changing FRD properties of the fibers) which otherwise could amount to several \si{\kmps}\footnote{Gabor Furesz, private communications.}. 

\begin{wrapfigure}[30]{r}{0.45\textwidth}
    \begin{minipage}{0.25\textwidth}
      \centering
        \includegraphics[trim={0cm 0cm 0cm 0cm}, clip,width=0.7\linewidth]{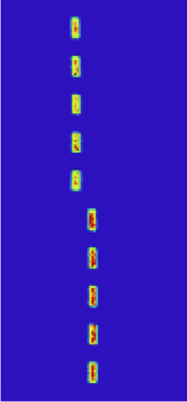}
    \end{minipage}%
    \begin{minipage}{0.2\textwidth}
      \caption{\textbf{Zemax image simulation of the five rectangular fibers forming the input slit} for two wavelengths (505.35 and 509.55\,nm) at the edge of the bluest orders (121 and 122). The order separation at this point is almost equal to the separation of the individual fibers.}
      \label{slitimage}
    \end{minipage}
\end{wrapfigure}
We only use a moderate anamorphism on the cross-disperser grisms of 1.23 to slightly compress the spectrum in spatial direction. The slit image is sampled by 3.5\,pixel per slit-width on average for both cameras (2.91\,pixel minimum). Due to the large separation of the individual fibers forming our slit (see Section\,\ref{fiber}), the separation of individual spectral traces in the same order (from the pupil slicer as well as from sky and calibration fibers) is \num{>10} pixels. The minimum inter-order separation is equally large (see Figure\,\ref{slitimage}).

\begin{figure}[!t]
\centering
\vspace{0mm}
\includegraphics[trim={0cm 2cm 0cm 1cm},width=1\linewidth,clip]{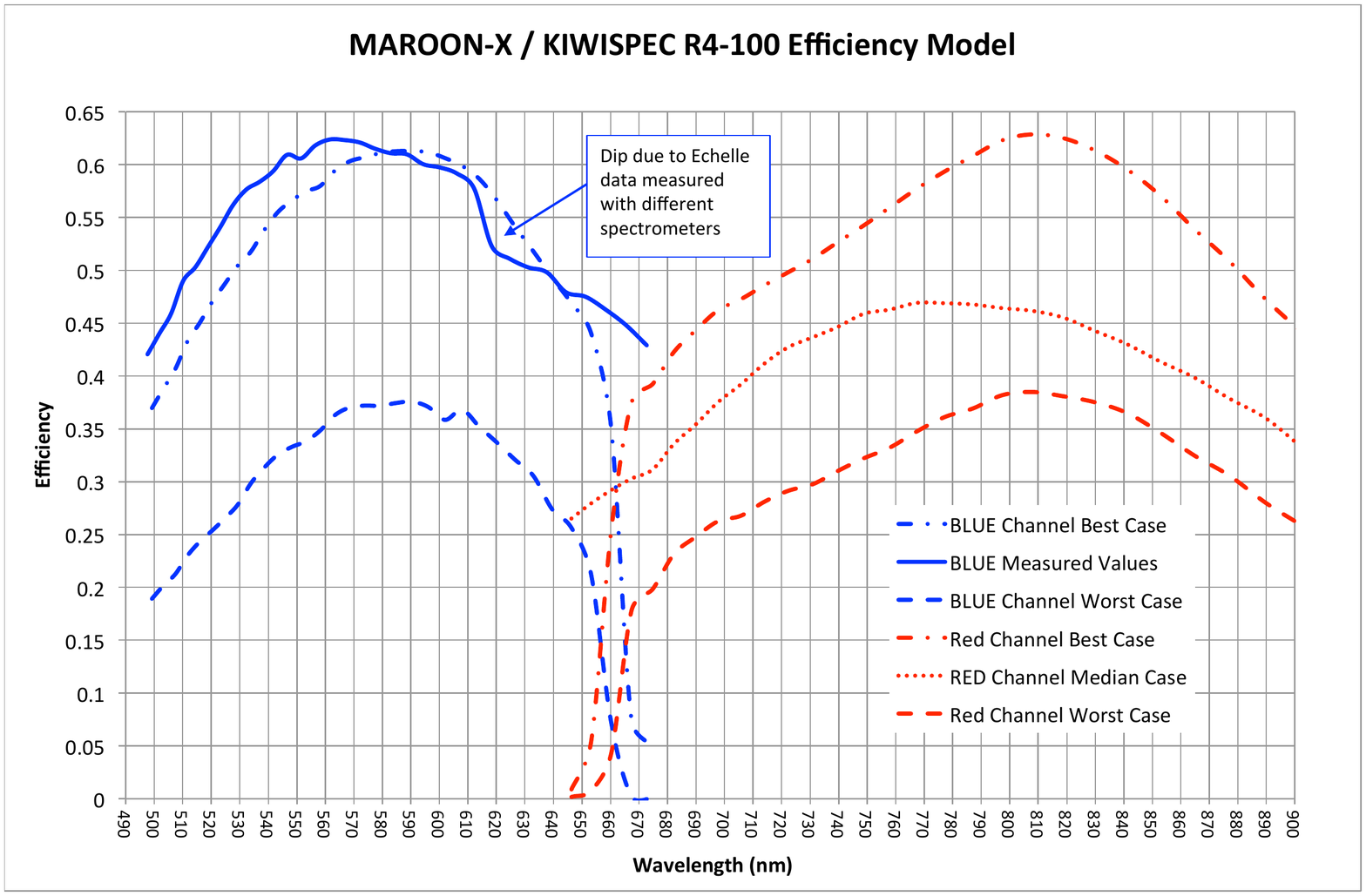}
\vspace{0mm}
\caption{\textbf{Instrument efficiency (intermediate slit to detector) for the blue and red arm, respectively}. Worst, median, and best estimates are based on vendor specifications for reflection and anti-reflection coatings, grating efficiencies, etc. Median cases pertain to average specifications, while worst and best cases pertain to minimum specs and goal values for all surfaces, respectively. The efficiency model is updated on the as-built version with measured efficiencies for the echelle grating, blue VPH grism and some of the blue camera optics (blue solid line) demonstrating the excellent throughput of the instrument. }
\label{efficiency}
\end{figure}

A pupil stop in the pre-optics and baffles around the beam footprint on the echelle grating and collimator mirror as well as in the camera barrel suppress unwanted stray light. Optical ghost analysis showed optical ghosts to be $<0.1\%$ of the intensity of the parent order for both cameras.
\clearpage
The current projections for the instrument efficiency are shown in Figure\,\ref{efficiency}. We expect peak throughputs for both arms to exceed 50\% (measured from $f/10$ slit image to detector plane). For the blue channel we already have as-built measured values for most of the critical items (echelle grating, cross-dispersers, mirror coatings) and find a peak throughput in excess of 60\%. 

For the exposure meter we collect part of the 0th order of the echelle grating with an off-the-shelf OAP (\SI{76}{\milli\meter} diameter, \SI{152}{\milli\meter} EFL) mounted behind the echelle. The OAP refocuses the un-dispersed image of the input slit onto a \SI{300}{\micrometer} diameter high-NA optical fiber which then guides the light outside the vacuum chamber. The size of the fiber is chosen to only accept the images of the three object fibers, rejecting the light from the sky and calibration fiber in order to avoid contamination of the spectro-photometric signal of the exposure meter.

\subsection{Mechanics}

A considerable effort went into numerical simulations of the spectrograph opto-mechanics to optimize thermo-mechanical stability vs. complexity and thus cost. These simulations have shown that a uniform application of stainless steel SS420 for all mounts and almost all structural components offers the best compromise between performance (thermal conductivity and CTE) and cost compared to other choices such as aluminum and Invar. We choose Invar only for the three legs of the internal table to minimize the impact of non-common expansion between the legs which would lead to beam steering on the cross-disperser and cameras exceeding our tolerances for RV stability.

The complete optical train from the input fibers up to and including the dichroic are enclosed in a vacuum chamber and will be held at $p$\,\SI{< e-3}{\milli\bar} and $dT$\,\SI{\leq 1}{\milli\kelvin} with a combination of active and passive temperature control. The cross-dispersers and camera optics are enclosed in a sealed barrel outside the vacuum chamber\footnote{We thank Francesco Pepe for this suggestion made during the PDR.}. This is permissible given their lower sensitivity to pressure and temperature changes compared to the echelle grating and minimizes the volume of the vacuum chamber. It also allows us to add the red wavelength arm and replace the detector modules at a later stage in the project without opening the vacuum chamber.

\begin{figure}[!t]
\centering
\vspace{0mm}
\includegraphics[trim={0cm 0cm 0cm 0cm},width=1\linewidth,clip]{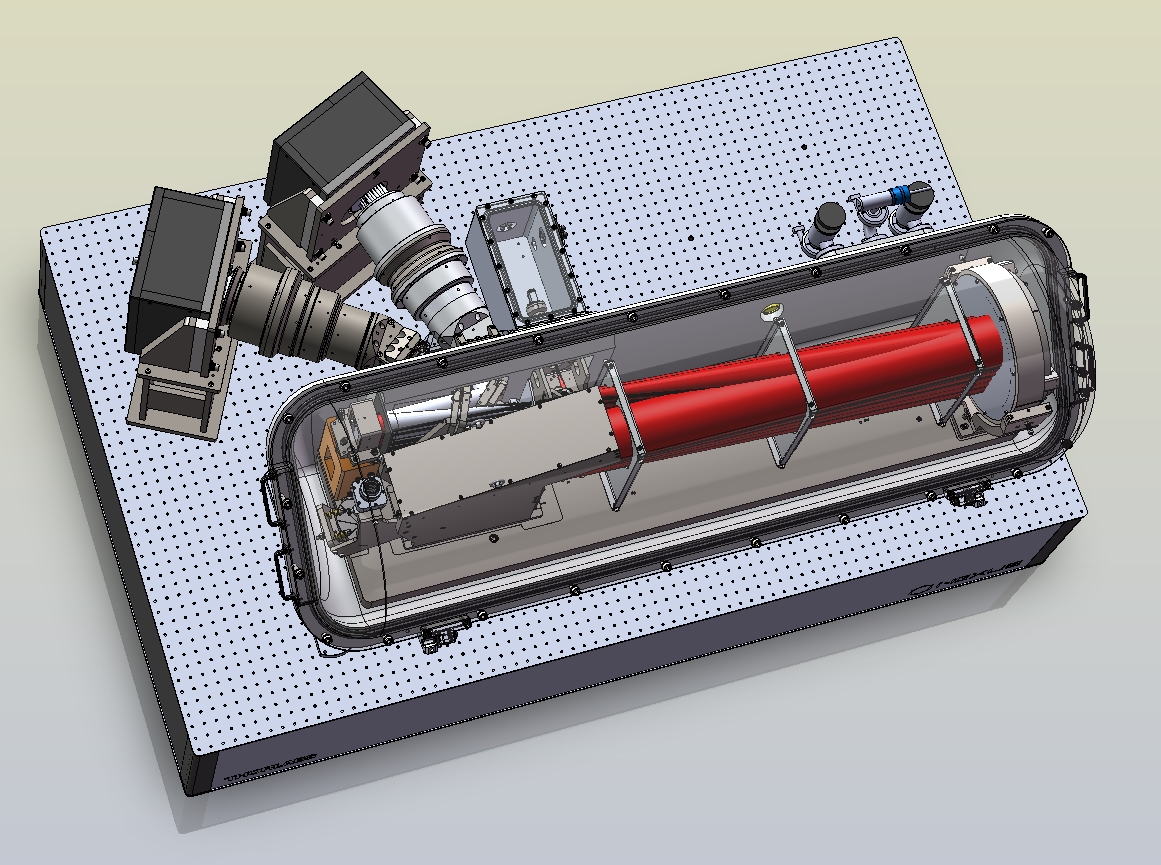}
\vspace{0mm}
\caption{\textbf{3D CAD rendering of KiwiSpec R4-100 for MAROON-X}. A transparent version of the top lid of the vacuum chamber reveals the main optical elements. The beam is shown in red and white. The two camera arms with small interim detector systems are shown to the top left. A small rectangular extension to the cylindrical vacuum chamber will house the pupil slicer. The pre-optics are mounted onto the internal bench but extend through the chamber wall into the extension chamber. The footprint of the spectrograph including the outside optical table is only \SI{1.2x2.1}{\meter}.} 
\label{CAD}
\end{figure}

We expect the pressure stability in the camera barrel to be still better than $dp$\,\SI{< e-1}{\milli\bar} and long term temperature drifts smaller than $dT$\,\SI{\leq 20}{\milli\kelvin}. The vacuum chamber is mechanically and thermally decoupled from the spectrograph optics. The optical components inside the vacuum chamber are mounted on an optical bench that rests on three Invar legs through vacuum bellow feedthroughs. The legs, as well as the separately mounted vacuum chamber, are bolted onto a commercial optical bench that sits on actively vibration dampened legs and is thermally insulated from the floor. There is no direct mechanical contact between the vacuum chamber and the optics (aside from the two vacuum windows) and a layer of PEEK between the top of the internal legs and the inner optical bench serves as thermal insulation. The vacuum chamber thus acts as a passive thermal insulation layer for the main optics.

Part of our characterization campaign in the lab, scheduled for the last quarter of 2016, is to determine the thermo-mechanical stability of the as-built spectrograph and implement modifications, if necessary. For example, the camera optics are not athermalized since we place them in a tightly thermally controlled environment. However, provisions are in place to add an athermalization mechanism to correct for focus and plate scale changes by changing the distance between the detector system and the camera barrel. Likewise, an additional passive or active insulation layer can be added to the vacuum vessel and the camera barrels to further dampen temperature fluctuations. 

The main characteristics of MAROON-X are summarized in Table\,\ref{table}.

\section{PUPIL SLICER \& IMAGE SCRAMBLER}\label{fiber}

KiwiSpec R4-100 would have a resolution-slit product of $R\phi\approx25,400$''  on the 6.5m Magellan Telescope. To achieve the desired resolving power of 80,000, we need to either slice the image or the pupil to reach an acceptable 
FOV on sky. Either technique will boost the efficiency of the spectrograph at the cost of an increase in slit height and thus spectral coverage and optical aberrations. 

We decided against image slicing, since throughput and alignment of image slicers are critical and the latter remains practically untested in the context of high-precision radial velocity work. It is important to realize that an image slicer effectively works as an anti-scrambler, as it non-linearly amplifies small changes in the fiber output illumination at the slicer edges. Moreover, being directly in the imaging plane of the spectrograph, a sub-m\,s$^{-1}$ stability requirement translates into nanometer levels of positional stability, which is challenging to control both on a mechanical and thermal level.

We thus decided to build a pupil slicer, which is much less critical in terms of (thermo-)mechanical stability. This technique has been selected for the next generation of large spectrographs (i.e. G-CLEF/GMT\cite{gclef,gclef2} and ESPRESSO/VLT\cite{espresso,espresso2}).

The median seeing on Cerro Manqui, the site of the twin Magellan Telescopes, is 0.62" and the 75-percentile is 0.79"\cite{thomas-osip}. We thus found a 3x slicer the best compromise between instrument throughput, allowable slit-height, and available fiber core sizes. With a \SI{100}{\micrometer} diameter octagonal fiber operated at $f/3.33$ we accept a 0.95" FOV on the $f/11$ focus of Magellan with a geometrical coupling efficiency between 81\% (under median seeing conditions) and 64\% (for the 75-percentile). 

Our design implementation of a pupil slicer uses microlens arrays (MLAs) to slice the pupil into three sections and to re-image the pupil slices onto three rectangular fibers with 1:3 aspect ratio. By feeding a pupil image into the slit-forming fibers, we are effectively incorporating a double scrambler into the pupil slicer as we use the octagonal fiber to scramble the stellar input image and the rectangular fibers to scramble the (sliced) pupil image. This also benefits the illumination stability of the slicer itself, as temporal instabilities in the slicing geometry will effectively be greatly reduced by the subsequent scrambling of the pupil images. 

The design parameters of the fibers and the slicer were again a compromise between a number of requirements and practical considerations, most notably to minimize focal-ratio-degradation (FRD) effects, achieve a good scrambling gain, keep the curvature of the microlenses within a manageable range, all the while working with non-custom fiber core geometries to avoid a costly custom fiber preform.

\begin{SCfigure}[][!t]
\centering
\vspace{0mm}
\includegraphics[trim={0cm 0cm 0cm 0cm},width=0.6\linewidth,clip]{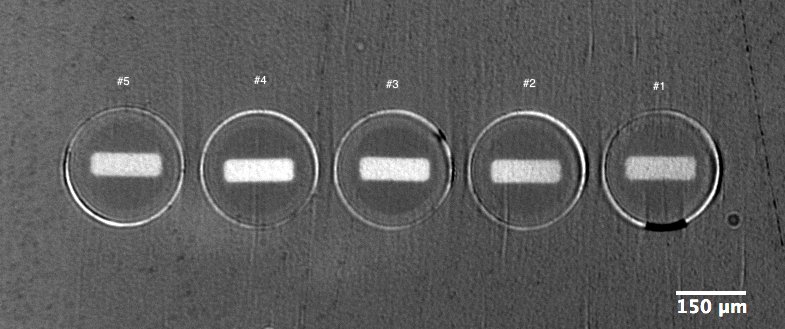}
\vspace{0mm}
\caption{\textbf{Prototype of the pseudo slit for MAROON-X}. Shown here is a FEMTOprint fiber slit plate with five Ceramoptec \SI{50x150}{\micrometer} rectangular fibers after polishing and before the wedge prism is glued on. 
}
\label{slit}
\end{SCfigure}

Our final design uses a \SI{100}{\micrometer} diameter octagonal fiber from CeramOptec (OCT-WF100/140/250, NA=0.22) at $f/3.33$ to guide the light from the telescope into the spectrograph through a custom vacuum feedthrough. We position all the optics of the pupil slicer inside the vacuum chamber. A small collimator forms a pupil image and two MLAs slice the telescope pupil into three sections and re-image these sections at $f/5$ with a separation of \SI{300}{\micrometer} onto three CeramOptec fibers with \SI{50x150}{\micrometer} rectangular cores. These fibers are positioned in a fused silica plate in a linear arrangement with the short sides of each fiber core lined up. After a fiber run of approximately \SI{1}{\meter}, in which the fibers are rotated by \ang{90}, another linear stack is formed, this time with the long sides of the fiber cores lined up to form the physical entrance slit of the spectrograph. At this point two more rectangular fibers are added to the three object fibers to add sky- and calibration light. An image of a slit plate prototype with five of the rectangular fibers forming a pseudo-slit is shown in Figure\,\ref{slit}. 

While technically within specification, the prototype slit plate has still sub-optimal alignment. The fibers were etched slightly too long, making them \SIrange{5}{6}{\micrometer} smaller than the holes in the plate, which leads to offsets. Likewise, rotational alignment of two fibers is off by \SI{\pm1.5}{\degree}. Achievable tolerances are \SIrange{2}{3}{\micrometer} in relative position and \SI{\pm0.3}{\degree} in rotation and the final slit plates will be produced to these specs.

A detailed description of the design, prototyping, and testing of our pupil slicer is presented elsewhere in these proceedings\cite{seifahrt}. Likewise, a discussion of the scrambling and FRD properties of our fibers can be found in another article in the same proceedings\cite{adam}.

Custom relay optics, consisting of a doublet and a triplet, form a 2x magnified, telecentric, image of the slit plate at the nominal $f/10$ focus of the spectrograph. The small wedge prism, part of the spectrograph camera optics, is directly bonded onto the fiber slit plate with Norland 61. The relay optics are fully diffraction limited over the full field and for all wavelengths. Typical rms radii range from \SIrange{1}{3}{\micrometer}.

The expected geometrical efficiency of the pupil slicer and double scrambler is limited by the achievable tolerances on the MLA optics and their alignment but is still $\geq$90\%. The pupil slicer and double scrambler unit has only four air-glass surfaces, three of which are BBAR coated. Five more air-glass surfaces, all of which are BBAR coated, are part of the spectrograph's pre-optics. Specifications on the BBAR coatings call for $R\leq0.5\%$ over 500--900\,nm. We thus estimate the efficiency of the optics that form the input slit of the spectrograph to be $\approx$78\%.

Additional losses from FRD effects are hard to fully quantify in advance, as they depend on a number of factors related mainly to the mechanical stress of the fibers at their mount points. We estimate 90\% here\cite{seifahrt}, which brings the combined efficiency of the light injection down to $\approx$70\%. Given the excellent throughput of the spectrograph itself, we still have enough margin left for the telescope front end in order to reach our combined throughput requirement of $\geq$11\%. 

\section{DETECTOR SYSTEM}\label{detector}

The final detector systems for MAROON-X will consist of two 4k$\times$4k \SI{15}{\micrometer} CCD detectors. For the red channel (650--900\,nm) we will use a deep-depletion, thick sensor to boost the quantum efficiency at the red end of the spectrum and to suppress fringing. We plan to use either two e2v CCD231-84 or two STA 4150 devices and a STA Archon readout system.

During assembly and initial lab testing we will use a commercial off-the-shelf CCD camera, a ProLine PL230 from Finger Lakes Instrumentation (FLI). This detector system is based on an e2v CCD 230-42 with 2k$\times$2k \SI{15}{\micrometer} pixels. The camera uses a Peltier cooler and the chip can be operated at \SI{-25}{\celsius} with a chilled water circulator to remove the excess heat.

We have removed the standard window of the ProLine PL230 camera and replaced it with a smaller version of our field-flattener lens, using a modified window holder. The inner chamber housing the CCD was pumped out for 48\,hrs. After subsequent back-filling with pure argon to atmospheric pressure, we re-sealed the CCD chamber. The expected thermo-mechanical stability and the performance of the readout system of our interim detector system will likely limit the achievable RV precision. However, this camera will still be useful for initial testing. 

\section{WAVELENGTH CALIBRATOR}\label{etalon}

Our main calibration system is a custom bulk Fabry-P\'erot etalon (FPI) from SLS with a finesse of 40 and a FSR of \SI{15}{\giga\hertz}. Illuminated by an Energetiq EQ-99XFC laser-driven light source (LDLS), the etalon produces a comb of unresolved lines across the entire bandpass of MAROON-X. 

The etalon is housed in a custom vacuum chamber with in-built liquid circulation channels by Lesker. We achieve vacuum levels of \SI{<e-6}{\milli\bar} with a \SI{0.5}{\litre\per\second} ion pump. A PolyScience PD07R-40 refrigerated circulator is used to stabilize the temperature of the etalon to \SI{<10}{\milli\kelvin} P-V. We monitor the temperature of the etalon with two PT100 sensors and a Lakeshore temperature controller. 

A grating stabilized external cavity diode laser (ECDL) from Toptica and a Rubidium cell from TEM Messtechnik is used to monitor one of the etalon lines for short- and long-term drifts. We scan the laser at a scan rate of \SI{4}{\hertz} over a spectral range of \SI{6.6}{\giga\hertz} (\SI{\approx5}{\kmps}) near 780.25\,nm to simultaneously measure one etalon line, three transition groups of Rb$^{85}$ and Rb$^{87}$ D2, and about 30 lines of a custom confocal etalon. The latter is used to linearize the scan axis and correct for piezo hysteresis effects. 

White light and laser light are combined and sent in a single mode fiber to the etalon. Vacuum compatible versions of Thorlab's off-axis paraboloid collimators are used to illuminate the etalon with collimated light. A \SI{50}{\micrometer} multi-mode fiber collects the light from the etalon and brings it out of the vacuum chamber.

An Ondax SureBlock ultra narrow-band notch filter is used to remove the laser line from the white light comb and reflect it onto a photo diode for measurement. The OD of the notch filter is about 4 and the FWHM is 0.45\,nm. This is sufficient to block the laser light and only lose etalon comb lines for about 5\% of the FSR in order 79 of the spectrograph. 

We achieve a rms frequency stability of the etalon of \SI{0.13}{\mega\hertz}, hence \SI{\sim10}{\cmps} within a few seconds. After binning 1\,min worth of data we reach a limit of \SI{3}{\cmps}. For longer timescales we find a linear drift of about \SI{30}{\cmps\per\day}, which we identify as the aging of the Zerodur spacer in the etalon, in full agreement to known shrinkage rates of Zerodur\cite{zerodur}. Thanks to the rubidium frequency standard, we can track the etalon zero point and either compensate or account for drifts and instabilities at the few \si{\cmps} level for timescales longer than 60\,s. 

A more detailed description of the design and performance of our etalon comb calibrator can be found elsewhere in these proceedings\cite{etalon}.

For in-depth characterization and initial testing of MAROON-X we will also use a classical ThAr lamp from Photron as well as an iodine absorption cell from Thorlabs. The ThAr lamp will provide a wavelength zero point as the etalon can only deliver relative positions until the order number of at least one etalon peak per echelle order is established and the true cavity length of the etalon is determined. The iodine cell will allow us to monitor the stability of the line spread function in the blue wavelength arm of the spectrograph, a useful tool for RV stability diagnostics.

\section{ENVIRONMENTAL CONTROL CHAMBER}\label{chamber}

In order to provide a tightly temperature controlled environment for the spectrograph, we installed a commercial walk-in freezer unit (Masterbilt 761012, \SI{10x12}{\foot}) in the lab. A dual-fan Lytron heat exchanger with \SI{2.7}{\kilo\watt} capacity and 700\,CFM air movement receives a temperature controlled water supply from a PolyScience PD15R-40 refrigerated circulator with \SI{1.1}{\kilo\watt} heating and cooling power. An external sensor at the output of the heat exchanger drives the PID loop of the circulator. Several temperature sensors, as well as pressure and humidity sensors inside and outside of the chamber allow a detailed monitoring of the conditions inside the chamber. 

We have not yet optimized the PID parameters and see air temperature fluctuations of \SI{30}{\milli\kelvin} P-V which even over weeks are fully dominated by 15\,min oscillations and clearly show the need for a re-tuning of the PID parameters. We currently run the chamber at a set-point \SI{5}{\celsius} lower than the average lab temperature, hence with a moderate but permanent forcing. 

Our lab environment is in two aspects similar to the foreseen location at the Magellan Telescope. The lab we placed the chamber in regularly sees temperature changes with up to \SI{2}{\celsius\per\hour} gradients due to malfunctioning HVAC supplies. These gradients are dampened to less than \SI{20}{\milli\kelvin} inside the chamber. The lab is in the sub-basement of a building and the temperature of the lab floor shows seasonal variations over timescales of months that change the floor temperature of the chamber by \SI{\approx200}{\milli\kelvin}. Only about \SI{20}{\milli\kelvin} of long-term temperature changes are recorded at a sensor farthest away from the heat exchanger. We thus clearly see the need of either actively controlling the floor temperate of the chamber or to thermally insulate the spectrograph from the chamber floor.

\begin{SCfigure}[][!t]
\centering
\vspace{0mm}
\includegraphics[trim={0cm 0cm 0cm 0cm},width=0.6\linewidth,clip]{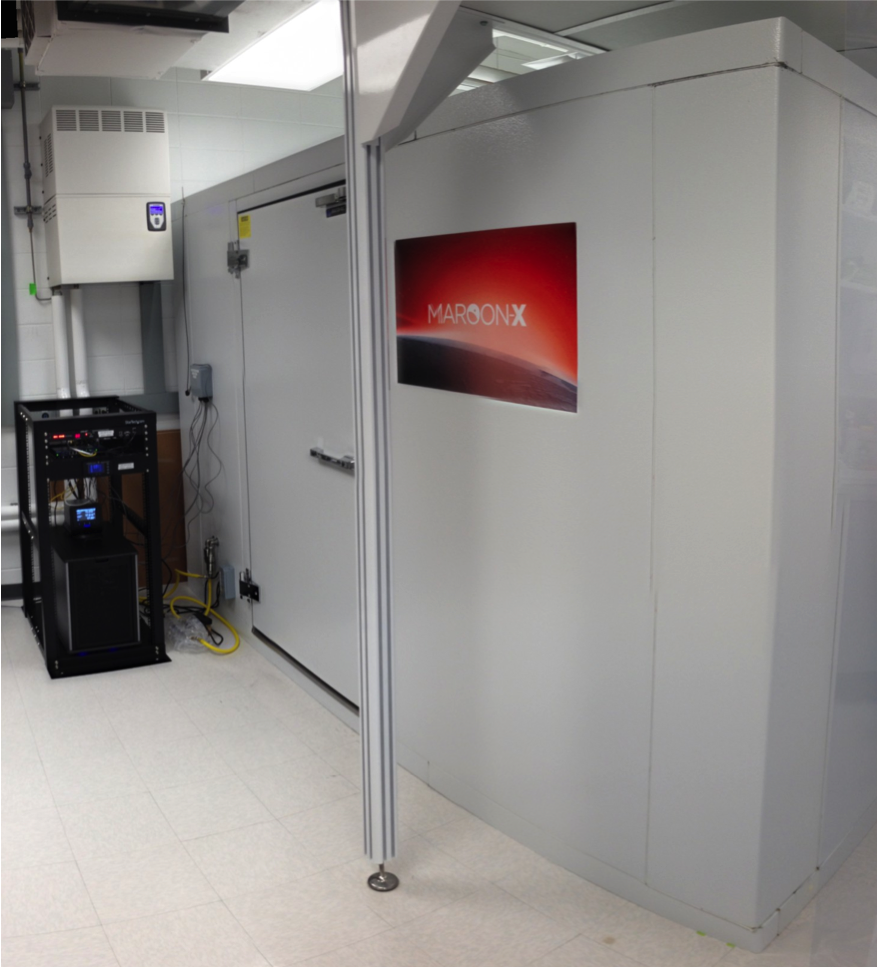}
\vspace{0mm}
\caption{\textbf{Environmental enclose in the lab in Chicago}. We currently achieve \SI{30}{\milli\kelvin} P-V inside the enclosure with a custom circulating bath and heat exchanger system. Enclosure floor and air temperatures are monitored at various locations. Pressure and humidity sensors provide complete environmental control.
}
\label{enclosure}
\end{SCfigure}

Before the arrival of the spectrograph we will tune the PID parameters of the circulator to improve our short-term temperature stability to better than \SI{5}{\milli\kelvin}. We will ultimately change the location of the temperature sensor for the PID loop to the location of the spectrograph to improve the long-term temperature stability under seasonal temperature forcings. We anticipate both long- and short-term stability of the air temperature surrounding the spectrograph of \SI{\leq10}{\milli\kelvin}.

All active (and hence heat dissipating) equipment (accept for the CCD detector systems) will be placed outside of the environmental chamber to minimize time-variable heat loads. The FLI camera, our interim detector system, is water cooled with a chiller and kept at a constant temperature. An additional layer of insulating foam panels is used to minimize the influence of the static temperature gradient between the CCD camera and the surrounding air. 

\begin{table}[!ht]
\caption{MAROON-X main characteristics}
\begin{center}
\begin{tabular}{|l|l|}
\hline
Spectral resolution & R = 80,000 for \SI{100}{\micrometer} slit image at $f/10$\\
Acceptance angle & FOV = 0.95'' at the Magellan 6.5m Telescope \\
Wavelength range & 500 nm -- 900 nm (in 56 orders)\\
Number and reach of arms & 2 (500--670\,nm and 650--900\,nm) \\
Cross-disperser & anamorphic VPH grisms\\
Beam diameter & 100\,mm (at echelle grating), 33\,mm (at cross-disperser)\\
Main fiber & \SI{100}{\micrometer} octagonal (CeramOptec)\\
Number and type of slicer & 3x pupil slicer \\ 
Slit forming fibers & 5$\times$ \SI{50x150}{\micrometer} rectangular (CeramOptec)\\
Inter-order and inter-slice spacing & $\geq10$ pixel \\
Average sampling & 3.5 pixel per FWHM\\
Blue detector & Standard 4k$\times$4k e2v or STA CCD (\SI{15}{\micrometer} pixel size)\\
Red detector & Deep-depletion 4k$\times$4k e2v or STA CCD (\SI{15}{\micrometer} pixel size)\\
Calibration & Fabry-P\'erot etalon for simultaneous reference (fed by 2nd fiber) \\
Environment for main optics & Vacuum operation, 1\,mK temperature stability\\
Environment for camera optics & Pressure sealed operation, 20\,mK temperature stability\\
Long-term instrument stability & \SI{0.7}{\mps} (requirement), \SI{0.5}{\mps} (goal)\\
Total efficiency & 11\% (requirement) to 15\% (goal) at \SI{700}{\nano\meter} (at median seeing)\\
Observational efficiency & S/N=100 at \SI{750}{\nano\meter} for a V=16.5 late M dwarf in \SI{30}{\minute} \\
\hline
\end{tabular}
\end{center}
\label{table}
\vspace{0mm}
\end{table}

\acknowledgments     
 
The University of Chicago group acknowledges funding for this project from the David and Lucile Packard Foundation through a fellowship to J.L.B.


\bibliography{report}   
\bibliographystyle{spiebib}   

\end{document}